\documentstyle[eqsecnum,aps,floats]{revtex}
\begin{document}
\tighten
\preprint{WATPHYS-TH98/01,gr-qc/yymmddd}

\title{Summary of Session A6, Alternative Theories of Gravity}

\author{Robert Mann}
\address{Physics Department}
\address{University of Waterloo, Waterloo, Ont. Canada N2L 3G1}

\date {March, 1998}

\maketitle

\section{Introduction}

The session on {\it Alternative Theories of Gravity} had perhaps the
greatest number of contributed papers out of all the parallel sessions
at GR15.  In all there were 84 contributions
which spanned an enormous breadth of alternative ideas concerning 
classical and quantum gravity.  For the most part, these ideas naturally 
grouped themselves into five categories: (a) Black Holes, (b) Tests of
Alternative Gravitational Theories, (c) Lower-Dimensional Gravity
(d) Membranes and Solitons and (e) Methods and Geometrical Structures.
There were 20 oral presentations made, with a total of four hours being
alloted to the sessions.  The sessions on black holes, tests and
lower dimensions were particularly well attended.  Below follows a brief
summary of each of the five sections. The sequence of oral presentations
was determined alphabetically (last name of first author) for each section.

\section{Black Holes}

There were three oral presentations given here. A. Bonanno presented some
recent results he obtained for mass inflation in higher derivative theories
of gravity \cite{Bonanno}.  The goal of this work is to gain some insight as to
how the classical description of the mass inflation singularity is altered
by quantum effects.  Such effects are generally modelled by modifying
the Einstein-Hilbert action so that it includes terms non-linear in the
curvature ({\it i.e.} the higher-derivative terms); these might be dynamically
generated during gravitational collapse. For the
Reissner-Nordstrom-Vaidya black holes, both the location and 
surface gravity of the inner horizon are modified. The impact of this
on the mass inflation phenomenon can only be determined by detailed
calculation, although it appears that the basic instability properties of
the Cauchy horizon will not change. 

The interior structure of non-Abelian black holes was the subject of
G. Lavrelashvili's talk \cite{GL}.  Unlike their Abelian counterparts,
no inner (Cauchy) horizon forms inside them. Instead, a different
sort of mass inflation occurs in which an enormous growth of the
mass function takes place just before a Cauchy horizon can form.
There are repeated cycles of this phenomenon in the absence of
a Higgs field, whereas no such cycles occur if a Higgs field is
present.

Finally, J. Koga (in collaboration with K. Maeda),
considered the behaviour of black hole thermodynamics in gravitational
theories in which the action is a functional of the metric, the Ricci
tensor, a scalar field and its derivative. These theories can be
converted to general relativity via a ``Legendre
transformation''.  For two black hole solutions
related by such a transformation, all thermodynamic variables 
are found to be the same, but the gravitational mass differs.

\section{Tests of Alternative Theories}

This part of the session had two talks. J. Novak spoke about the
possible empirical interest of tensor-scalar gravity: even if
it is tightly constrained by solar-system experiments to behave
almost like general relativity in the weak-field limit, it could
differ from it significantly in the strong-field regime. Compact
objects  ({\it e.g.} neutron stars) would radiate scalar gravitational 
waves as they collapsed, and these could be interferometrically
detected with technology currently under construction.
Results from numerical simulations of the collapse of a stellar core 
into a neutron star were presented. This work will soon be published \cite{JN}.

The other talk by A. Edery (collaborating with M. B. Paranjape), 
considered the empirical viability of Weyl gravity
in the context of light deflection of light \cite{AE}. 
An extra deflection term beyond
the usual one in general relativity is obtained, and this term
is significant at large (galactic or greater) distance scales where dark matter
effects are observed.
However the sign of the extra correction term is opposite to that
needed to fit galactic rotation curves.  This leads to two possibilities:
either Weyl gravity cannot solve the dark matter problem or fitting galactic 
rotation curves in a conformal theory(a massless theory) is in the first 
place not a reliable procedure until one introduces a symmetry 
breaking mechanism.

\section{Lower-Dimensional Gravity}

This was quite a popular section of the session, with a considerable number
of oral and poster presentations. M. Cataldo \cite{MC} (collaborating with
P. Salgado) described work which shows how to obtain a solution to the
$(2+1)$-dimensional Einstein-Maxwell equations for a charged spinning
black hole, correcting previous work by Kamata and Koikawa on this problem.
Furthermore, a electromagnetic duality mapping amongst different spinning 
solutions was shown to exist.

R. Tavakol (working with J.E. Lidsey and C. Romero) presented work which
described a constructive method for finding higher-dimensional vaccum
solutions from those in lower-dimensions by making use of an
embedding theorem due to Campbell. Examples include 
finding local embeddings of general relativistic solutions
in Ricci-flat 5-dimensional spaces as well as showing how to
relate $(2+1)$--dimensional gravity to vacuum  $(3+1)-$dimensional 
general relativity. Tavakol also claimed that this approach could
yield new solutions.  This work has recently been published \cite{RT}

T. Ohta (in collaboration with R.B. Mann) described some preliminary
results of a consideration of the 2-body problem in $(2+1)$-dimensional
general relativity. An exact solution for the reduced Hamiltonian as 
a function of the relative position and momentum can be obtained in 
the case of two massless bodies.  This solution furthermore has a very
interesting resemblance to an exact solution to the 2-body problem
they recently obtained in $(1+1)$ dimensions \cite{TO}.  Further implications of
these results are being worked out.

B. Paul (working with S. Mukherjee and A. Beesham), interested in
investigating dissipative thermodynamic effects in cosmological models,
described results of an investigation of such effects in $(1+1)$-dimensional
gravity, where mathematical simplicity affords one considerable
computational progress.  $R=T$ theory was chosen because a considerable number
of its properties and solutions closely resemble those of general relativity.
A number of results were obtained, including a demonstration of the
unphysical aspects of Eckart theory, solutions with oscillating Hubble
parameter, and double-inflation solutions.

J. Soda gave a talk showing how 2-dimensional gravity (including constant
curvature theories, the CGHS model, and spherically symmetric gravity) 
could be pressed into service to shed light on critical phenomena 
in gravitational collapse. Self-similar solutions are generally believed to
provide a good approximation to the non-linear dynamics present in 
gravitational collapse. However Soda finds that not all self-similar models 
elicit critical phenomena. 

Finally, Vendrell gave a talk described a new black-hole solution
of $(1+1)$-dimensional ``$R=T$'' theory which is the
endpoint of the collapse of an infinitesimally thin shell of 
radiation. The interior singularity has the topology of a corner,
and the black hole may be considered as a
two-dimensional analog of the Schwarzschild black hole.
Vendrell discussed some of its semi-classical and quantum properties.
Some of this work has been published \cite{FV}.

\section{Membranes and Solitons}

This part of the alternative theories session dealt with new ideas
for incorporating extended objects into gravitational theory.
D. Gal'tsov (in collaboration with Chen) described work in which
sigma--model representations were derived for $P$--branes
corresponding differing decompositions of the full metric. 
Transformations which generate solutions preserving asymptotic flatness 
of the metric were discussed, and in the case where the solutions
depended on  two transverse coordinates, associated Geroch--type symmetries
are present.

$P$-branes were employed in a somewhat different way in talk
given by M. Grebniuk (with V.D. Ivashchuk V.D. and V.N. Melnikov).
He began with a multidimensional cosmological model that described 
the behaviour of $n+1$ Einstein spaces for which the action contained 
a number of dilatonic scalar fields $\varphi^a$ and antisymmetric 
forms $A^I$. The $P$-branes enter when the $\varphi^a$ forms
are chosen to be proportional to the volume elements of the "$p$-brane" 
submanifolds. The general motivation behind this work appears to
be in understanding the general ways in which gravity (in a cosmological
setting) might be ultimately described in terms of some kind of membrane
physics. A Toda-like Lagrangian representation arises, and it is
possible to obtain exact solutions to the field equations in
certain circumstances. 

M. Pavsic gave a talk in which he proposed that spacetime itself be considered as
an $n$-dimensional membrane embedded in a (presumably extra-physical) $N$ dimensional
manifold. As no constraints are imposed on the membrane, both its 
normal and tangential motions are physically relvant, and 
the theory can be straightforwardly quantized. The distinction between the evolution 
parameter of the membrane's wavepacket is distinct from the timelike coordinate
of the membrane, thereby offering an approach for resolving the
"problem of time" in quantum gravity.

The last talk of this section was by M. Sethi (working with D. Lohiya).
In a Brans--Dicke cosmology where the scale factor evolves linearly 
with time, non--topological soliton solutions can arise in which
spacetime breaks up into domains with differing values of the
effective gravitational coupling constant. A domain consisting
of a region having the canonical value of the gravitaional constant
surrounded by one in which this constant vanishes is called
a gravity ball (or (g)--ball). In this toy model, 
gravitating matter is trapped inside gravity(g)--balls as 
large as (say) the halo of typical galaxies. The authors claim
that the resulting cosmology has no horizon, flatness or
cosmological constant problems and that it is consistent with the three
"classical tests" of cosmology: Number Count--Redshift, Luminosity
 distance--Redshift and the Angular Diameter--Redshift. Some results of
this work are available in preprint form \cite{MS}.

\section{Geometrical Structures}

This section of the session was devoted to a consideration of ideas which
question the basic geometric notions which undergird general relativity and
which seek alternative means of understanding them. Approximately 30 papers
contributed to the session were in this category, the largest out of all the
sections. There were five oral presentations given.

L. Querella (working with S. Cotsakis and J. Miritzis) described the results
of an investigation \cite{LQ} of a metric-affine approach to  higher order theories 
of gravity in which the Lagrangian is an arbitrary function of the curvature 
invariants.  A conformal equivalence theorem between these theories and general relativity 
plus a scalar field was shown to hold in the extended framework of 
Weyl geometry with the same forms of field and self-interacting potential. 
However a new `source term' appears which makes an additional contribution
to the stress-energy. This approach  may lead consistently to 
a new method of reduction of order of the associated field equations.

M. Mars (collaborating with R.M. Zalaletdinov) gave a talk in which
he described a new approach for covariant space-time averaging of
tensorial quantities on differentiable metric manifolds with a
volume $n$-form. A new result was presented
demonstrating that the averaging bilocal operator is idempotent iff 
it is factorized into a bilocal product of a matrix-valued function 
on the manifold, taken at a point, by its inverse at another point.
Several other new results concerning the algebraic structure of
the averaging operator were also mentioned.

E. Poberii's contribution presented us with the suggestion
that general relativity be considered as the low-energy limit
of a theory in which metric compatibility is not required,
torsion is permitted, and local conformal invariance holds.
Such a theory is postulated to be relevant in the very early
universe.  The dynamical non-metricity present in the theory
can induce a type of chaotic inflation. During inflation the
non-metricity tends to zero, so that our space-time currently 
appears to be Riemannian with a high degree of accuracy.

Non-metricity also played an important role in U. Schelb's talk.
He claimed that non-metricity could provide a mathematical criterion 
referring to non-spacelike geodesics for 
distinguishing between Riemannian and Weylian geometries. This in
turn could be translated into operationally testable
criteria based on distance measurements with light signals,
without any recourse to a second clock effect 
in Weyl space or to atomic clocks.

The final talk of the session was given by P. Smrz, who described
an approach toward describing physical fields in geometrical terms
that he has been exploring for a number of years \cite{PS}.
In this approach the classical properties of space-time are 
derived from the geometry of a four-dimensional complex manifold with 
a linear connection. The measurement process forces the properties of space-time 
to depend on the choice of the reference cross section. The
present talk described how one might hope to provide a geometric 
explanation for fermions.

\section{Closing Remarks}

The large number of abstracts submitted to this session 
is indicative of the robust level of interest members of the theoretical
physics and mathematics communities express in considering
alternatives to general relativity.  Although several of these alternatives
are very speculative, an increasing number of them are plausible variants
of some of the basic ideas that underlie our understanding of gravitational
theory.  Their chief (and often under-appreciated) function is that they
continually remind us that there are fresh perspectives on old (and sometimes
not-so-old) problems.


\begin{references}{}


\bibitem{Bonanno}A. Bonanno ``The Cauchy Horizon in Higher-derivative Gravity Theories'',
gr-qc/9801077.

\bibitem{GL}P. Breitenlohner, G. Lavrelashvili, D. Maison
``Mass inflation inside non-Abelian black holes'', gr-qc/9711024. 

\bibitem{JN}J. Novak,
``Spherical neutron star collapse toward a black hole in 
tensor-scalar theory of gravity'', gr-qc/9707041 (to appear in Phys. Rev. {\bf D}).

\bibitem{AE}A. Edery and M.B. Paranjape,
``Classical tests for Weyl gravity: deflection of light and radar echo delay''
 astro-ph/9708233.

\bibitem{MC}M. Cataldo and P. Salgado, 
``The Electrically Charged Extreme BTZ Black Hole 
with Self (Anti-self) Dual Maxwell Field'', gr-qc/9611004. 

\bibitem{RT}J. Lidsey, C. Romero, R. Tavakol, and
S. Rippl,  Class. Quantum Grav., {\bf 14}, (1997) 865.

\bibitem{TO}T. Ohta and R.B. Mann, Class. Quant. Grav. {\bf 14} (1997) 1259.

\bibitem{JS}J. Soda and  K. Hirata,  Phys. Lett. {\bf B387} (1996) 271.

\bibitem{FV}F. Vendrell, ``A black hole in two-dimensional space-time'',
hep-th/9705135; Helv. Phys. Acta {\bf 70} (1997) 598.

\bibitem{MP}M. Pavsic, Nuovo Cim. {\bf 110A} (1997) 369;  Nucl. Phys. Proc. Suppl.
{\bf 57} (1997) 265; Found Phys. {\bf 26} (1996) 159; 
Found. Phys. {\bf 25} (1995) 819.

\bibitem{MS}M. Sethi {\it et. al.}, ``A Program for a Problem Free Cosmology Within 
a Framework of a Rich Class of Scalar-tensor Theories''; IUCAA-61-97, (Dec 1997). 

\bibitem{LQ}S. Cotsakis, J. Miritzis and L. Querella, ``Variational and Conformal 
Structure of Nonlinear Metric-connection Gravitational Lagrangians'', gr-qc/9712025.

\bibitem{US}U. Schelb, Int. Journ. Theor. Phys. {\bf 35} (1996) 1767;
 {\bf 36} (1997) 1341.

\bibitem{PS}P.K.Smrz, Aust. J. Phys. {\bf 50} (1997) 793; Aust. J. Phys. {\bf 48} 
(1995) 1045; J. Math. Phys. {\bf 28} (1987) 2824. 


\end{references}
\end{document}